\documentclass[fleqn]{ckm}                 % twocolumn proceedings style
\usepackage{txfonts}            % if you have it, to get Times family
\usepackage{epsf}

\confname{Workshop on the CKM Unitarity Triangle, IPPP Durham, April
  2003}

\title{Weak Decay Form Factors from QCD Sum Rules on the Light-Cone}

\author{Patricia Ball}
\address{IPPP, University of Durham}

\begin{document}

\begin{abstract}
I present a compilation of results on $B\to\,$light meson form factors
from QCD sum rules on the light-cone.
\end{abstract}

%% \maketitle needs to be after the author and address info and the
%% abstract... 
\maketitle

%% standard LaTeX from here on...

Form factors of B to light meson transitions are not only needed for
the extraction of $|V_{ub}|$ from $B\to\pi e \nu$ or $B\to\rho e \nu$
and $|V_{ts}|$ and $|V_{td}|$ from $B\to K^*\gamma$ and
$B\to\rho\gamma$, respectively (provided there is no new physics in
these decays), but they also enter most prominently the calculation of
$B\to\,$charmless nonleptonic decays in BBNS factorisation \cite{BBNS}.
It is hence of eminent importance to calculate them as precisely as
possible. 
The most precise calculation of the form factors
will undoubtedly finally come from lattice simulations; the present
state of this art is summarised in \cite{onogi}.
Another, technically much simpler, but also 
less rigorous approach is provided by QCD sum rules on the
light-cone (LCSRs) \cite{BBK,CZ}. The key idea is to
consider a correlation function of the weak current and a current with
the quantum-numbers of the B meson, sandwiched between the vacuum and
the meson $M$, i.e.\ $\pi$, $K$, $\eta$, $\eta'$, $\rho$, $\omega$, $K^*$
or $\Phi$. For large (negative) virtualities of these currents, the
correlation function is, in coordinate-space, dominated by distances
close to the light-cone and can be discussed in the framework of
light-cone expansion. In contrast to the short-distance expansion
employed by conventional QCD sum rules \`a la SVZ \cite{SVZ}, where
nonperturbative effects are encoded in vacuum expectation values 
of local operators with
vacuum quantum numbers, the condensates, LCSRs
rely on the factorisation of the underlying correlation function into
genuinely nonperturbative and universal hadron distribution amplitudes (DAs)
$\phi$ that are convoluted with process-dependent amplitudes $T_H$,
which are the analogues to the Wilson-coefficients in the
short-distance expansion and can be
calculated in perturbation theory, schematically
\begin{equation}\label{eq:1}
\mbox{correlation function~}\sim \sum_n T_H^{(n)}\otimes \phi^{(n)}.
\end{equation}
The sum runs over contributions with increasing twist, labelled by
$n$, which are suppressed by
increasing powers of, roughly speaking, the virtualities of the
involved currents. 
The same correlation function can, on the other hand, be written as a
dispersion-relation, in the virtuality of the current coupling to the
B meson. Equating dispersion-representation and the
light-cone expansion, and separating the B meson contribution from
that of higher one- and multi-particle states, one obtains a relation
(QCD sum rule) for the form factor describing the $B\to M$ transition. 

The particular strength of LCSRs lies in the
fact that they allow inclusion not only
of hard-gluon exchange contributions, which have been identified, in
the seminal papers that opened the study of hard exclusive processes
in the framework of perturbative QCD (pQCD)
\cite{pQCD}, as being dominant in light-meson form
factors, but that they also capture the so-called
Feynman-mechanism, where the quark created at the weak vertex carries
nearly all momentum of the meson in the final state, while
all other quarks are soft. This mechanism is suppressed by two powers
of momentum-transfer in processes with light mesons; as shown in
\cite{CZ}, this suppression is absent in heavy-to-light
transitions\footnote{For very large quark masses, though, the
  Feynman-mechanism is suppressed by Sudakov-logarithms, which
  are, however, not expected to be effective at the b quark
  mass, cf.\ for instance \cite{descotes}.} and hence any reasonable 
application of pQCD to B meson
decays should include this mechanism. LCSRs also avoid any reference to a
``light-cone wave-function of the B meson'', which is a necessary
ingredient in all extensions of the original pQCD method
to heavy-meson decays
\cite{BBNS,the_hard_guys}, including factorisation formulas obtained in 
SCET \cite{SCET}, 
but about which only very little is known.
A more detailed discussion of the
rationale of LCSRs and of the more
technical aspects of the method can 
be found e.g.\ in \cite{LCSRs:reviews}.

LCSRs are available for the $B\to\pi,K$ form
factor $f_+$ to
$O(\alpha_s)$ accuracy for the twist-2 and part of the twist-3
contributions
and at
tree-level for higher-twist (3 and 4) contributions
\cite{Bpi,solo,zwicky}. 
For the $B\to\,$vector transitions, the sum rules are known to 
$O(\alpha_s)$ accuracy for the twist-2 contributions and at tree-level for 
twist-3 and 4 contributions \cite{Brho,BB98}; ditto for $B\to\gamma$ 
\cite{Kou}.

Let us now properly define the form factors in question.
For a pseudoscalar meson $P$ we have  
($q=p_B-p$)
\begin{eqnarray}
\lefteqn{\langle P(p) | \bar q \gamma_\mu b | B(p_B)\rangle =  
f_+(q^2) \left\{(p_B+p)_\mu - \frac{m_B^2-m_P^2}{q^2} \, q_\mu 
\right\}}\nonumber\\
&&{} +
\frac{m_B^2-m_P^2}{q^2} \, f_0(q^2)\, q_\mu,\label{FF1}\\
\lefteqn{\langle P(p) | \bar q \sigma_{\mu\nu} q^\nu (1+\gamma_5) b | B(p_B)
\rangle }\nonumber\\
&=&i\left\{ (p_B+p)_\mu q^2 - q_\mu (m_B^2-m_P^2)\right\} \,
  \frac{f_T(q^2)}{m_B+m_P},
\end{eqnarray}
whereas for a vector meson $V$ with polarisation vector $\epsilon_\mu$:
\begin{eqnarray}
\lefteqn{
\langle V(p) | \bar q \gamma_\mu (1-\gamma_5) b | B(p_B)\rangle  =  
\epsilon_{\mu\nu\rho\sigma}\epsilon^{*\nu} p_B^\rho p^\sigma\,
\frac{2V(q^2)}{m_B+m_V}}\\
&&{} -i \epsilon^*_\mu (m_B+m_V)
A_1(q^2)+ i (p_B+ p)_\mu (\epsilon^* p_B)\,
\frac{A_2(q^2)}{m_B+m_V}\\
& & {}+i
q_\mu (\epsilon^* p_B) \,\frac{2m_V}{q^2}\,
\left(A_3(q^2)-A_0(q^2)\right)\\
&&\mbox{with~~} A_3(q^2) = \frac{m_B+m_V}{2m_V}\, A_1(q^2) -
\frac{m_B-m_V}{2m_V}\, A_2(q^2),\nonumber
\end{eqnarray}
$A_0(0) = A_3(0)$ and $\langle V |\partial_\mu A^\mu | B\rangle  =  2 m_V
(\epsilon^* p_B) A_0(q^2)$, and
\begin{eqnarray}
\lefteqn{\langle V(p) | \bar q \sigma_{\mu\nu} q^\nu (1+\gamma_5) b |
B(p_B)\rangle = i\epsilon_{\mu\nu\rho\sigma} \epsilon^{*\nu}
p_B^\rho p^\sigma \, 2 T_1(q^2)}\\
& & {} + T_2(q^2) \left\{ \epsilon^*_\mu
  (m_B^2-m_{K^*}^2) - (\epsilon^*p_B) \,(p_B+p)_\mu \right\}\\
& & {} + T_3(q^2)
(\epsilon^*p_B) \left\{ q_\mu - \frac{q^2}{m_B^2-m_{K^*}^2}\, (p_B+p)_\mu
\right\}\label{FF2}
\end{eqnarray}
with $T_1(0) = T_2(0)$. In semileptonic decays the 
physical range in $q^2$ is $0\leq q^2\leq (m_B-m_{P,V})^2$, which can
reach values up to $\sim25\,\mbox{GeV}^2$.

The starting point for the calculation of e.g.\ the form factor
$f_+$ for $B\to\pi$  is the
correlation function
\begin{eqnarray}
\lefteqn{i\int d^4y e^{iqy} \langle \pi(p)|T[\bar q\gamma_\mu b](y)
[m_b\bar b i\gamma_5 q](0)|0\rangle}\\
&&=\Pi_+ 2p_\mu + \dots,\label{eq:CF}
\end{eqnarray}
where the dots stand for structures not relevant for the calculation
of $f_+$. For a certain configuration of
virtualities, namely $m_b^2-p_B^2\geq O(\Lambda_{\rm QCD}m_b)$ and 
$m_b^2-q^2\geq
O(\Lambda_{\rm QCD}m_b)$, the integral is dominated by light-like distances 
and accessible to an expansion around the light-cone:
\begin{equation}\label{eq:3}
\Pi_+ (q^2,p_B^2) = \sum_n \int_0^1 du\, \phi^{(n)}(u;\mu_{\rm F}) 
T_H^{(n)}(u;q^2,p_B^2;\mu_{\rm F}).
\end{equation}
As in (\ref{eq:1}), $n$ labels the twist of operators and 
$\mu_{\rm F}$ denotes the factorisation scale. The restriction
on $q^2$, $m_b^2-q^2\geq O(\Lambda_{\rm QCD}m_b)$, 
implies that $f_+$ is not accessible at all momentum-transfers; to
be specific, we restrict ourselves to $0\leq q^2\leq 14\,$GeV$^2$.
As $\Pi_+$ is independent
of $\mu_{\rm F}$, the above formula implies that the scale-dependence of
$T_H^{(n)}$ must be canceled by that of the DAs $\phi^{(n)}$. 

In (\ref{eq:3}) we have assumed that $\Pi_+$ can be described by
collinear factorisation, i.e.\ that the only relevant degrees of
freedom are the longitudinal momentum fractions $u$ carried by the
partons in the $\pi$, and that
transverse momenta can be integrated over. Hard infrared (collinear) 
divergences occurring in $T_H^{(n)}$ should be absorbable into the
DAs. Collinear factorisation is trivial at tree-level,
where the b quark mass acts effectively as regulator,
but can, in principle, be violated by radiative corrections, by
so-called ``soft'' divergent terms, which yield divergences upon
integration over $u$. Such terms break, for instance, factorisation in
non-leading twist in the treatment of nonleptonic B decays \`a la BBNS
\cite{BBNS}. For the simpler case of the correlation function
(\ref{eq:CF}), on the other hand, where
the convolution involves only one DA instead of up
to three in $B\to\pi\pi$, it was shown in \cite{zwicky} that
factorisation also works at one-loop level for twist-3 contributions
and that there are no soft divergences.

As for the distribution amplitudes (DAs), they have been discussed
intensively in the literature, cf.\ \cite{DAs,BKM}. For instance, to 
leading order in
the twist expansion, there are three DAs for light mesons, which are
defined by the following light-cone matrix elements ($x^2=0$):
\addtolength{\arraycolsep}{-3pt}
\begin{eqnarray*}
\langle 0 | \bar u(x)\gamma_\mu\gamma_5 d(-x) |
 P(p)\rangle & = & i
f_P p_\mu \int_0^1 du e^{i\xi px} \phi_P(u),\\
\langle 0 | \bar u(x) \gamma_\mu d(-x) | V(p)\rangle & = &
f_V m_V p_\mu \,\frac{\epsilon x}{p x}\,
\int_0^1 \!\! du \,e^{i \xi px} \,\phi_\parallel(u),\\
\langle 0 | \bar u(x) \sigma_{\mu\nu} d(-x) | V(p)\rangle &=&\\
\lefteqn{=
i f_V^T(\mu) (\epsilon_\mu p_\nu - p_\mu \epsilon_\nu)
\int_0^1 \!\! du \,e^{i\xi px} \,\phi_\perp(u),}\hspace*{1.5cm}&&
\end{eqnarray*}
where $\xi=2u-1$ and we have suppressed the Wilson-line $[x,-x]$
 needed to ensure
gauge-invariance. 
The sum rule calculations performed in \cite{Bpi,solo,zwicky,Brho,BB98} include
all contributions from DAs up to twist-4. The DAs are parametrized by
 their partial wave expansion in conformal spin, which to NLO provides a
 controlled and economic expansion in terms of only a few hadronic
 parameters, cf.\ \cite{DAs} for details.

Let us now derive the LCSR for $f_+$. The
correlation function $\Pi_+$, calculated for unphysical
$p_B^2$, can be written as dispersion relation over its physical cut. Singling
out the contribution of the B meson, one has
\begin{equation}\label{eq:corr}
\Pi_+ =  f_+(q^2) \, \frac{m_B^2f_B}{m_B^2-p_B^2}
+ \mbox{\rm higher poles and cuts},
\end{equation}
where $f_B$ is the leptonic decay constant of the B meson,
$f_Bm_B^2=m_b\langle B| \bar b i\gamma_5 d|0\rangle$.
In the framework of LCSRs one does not use (\ref{eq:corr}) as it stands,
but performs a  Borel transformation,
$1/(t-p_B^2)\to \hat{B}\, 1/(t-p_B^2) = 1/M^2 \exp(-t/M^2)$,
with the Borel parameter $M^2$; this transformation enhances the
ground-state B meson contribution to the dispersion representation of $\Pi_+$
and suppresses contributions of higher twist to the light-cone expansion of
$\Pi_+$. The next step is to invoke quark-hadron
duality to approximate the contributions of hadrons other than the
ground-state B meson by the imaginary part of the light-cone
expansion of $\Pi_+$, so that
\begin{eqnarray}
\hat{B}{\Pi_+^{\rm LCE}} & = &
\frac{1}{M^2}\, m_B^2f_B \,f_+(q^2)\,e^{-m_B^2/M^2}\nonumber\\
&&{} +
\frac{1}{M^2}\, \frac{1}{\pi}\int_{s_0}^\infty \!\! dt \, {\rm
Im}{\Pi^{\rm LCE}_+}(t) \, \exp(-t/M^2)\nonumber\\
{\rm and}\quad \hat{B}_{\rm sub}\Pi_+^{\rm LCE} & = & \frac{1}{M^2}\,
m_B^2f_B \,f_+(q^2)\,e^{-m_B^2/M^2}.\label{eq:SR}
\end{eqnarray}
Eq.~(\ref{eq:SR}) is the LCSR for $f_+$.
$s_0$ is the so-called continuum
threshold, which separates the ground-state from the continuum
contribution. At tree-level, the continuum-subtraction in
(\ref{eq:SR}) introduces a lower limit of integration, $u\geq
(m_b^2-q^2)/(s_0-q^2)\equiv u_0$, in (\ref{eq:3}), which behaves as
$1-\Lambda_{\rm QCD}/m_b$ for
large $m_b$ and thus corresponds to the dynamical
configuration of the Feynman-mechanism, as it cuts off low momenta of
the u quark created at the weak vertex. At $O(\alpha_s)$, there are
also contributions with no cut in the integration over $u$, which
correspond to hard-gluon exchange contributions. Numerically, these
terms turn out to be very small, $\sim O(1\%)$ of the total result for $f_+$.
As with standard QCD sum rules, the use of quark-hadron
duality above $s_0$ and the
choice of $s_0$ itself introduce a certain model-dependence (or
systematic error) in the final result for the form factor, which is
difficult to estimate. To be on the conservative side,
one usually adds a 10\% systematic error to the final result for $f_+$. 
\begin{figure}[tbh]
\centerline{\epsfxsize=0.4\textwidth\epsffile{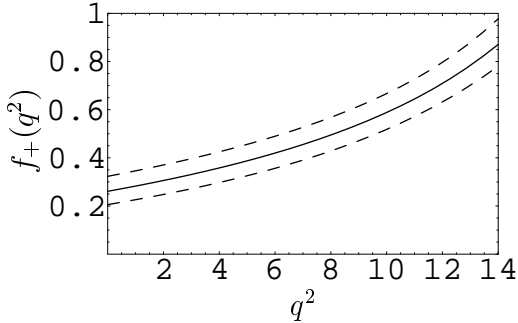}}
\caption[]{The $B\to\pi$ form factor $f_+(q^2)$ as function of 
momentum-transfer $q^2$ from the LCSR
  (\protect{\ref{eq:SR}}). Solid line: LCSR for central values of
input parameters and $M^2=6\,$GeV$^2$. Dashed lines: dependence of $f_+$ on
variation of input parameters and
$5\,$GeV$^2\leq M^2 \leq 8\,$GeV$^2$. Figure taken from 
Ref.~\cite{zwicky}.}\label{fig:1}
\end{figure}

Putting everything together, we obtain $f_+(q^2)$ as plotted in
Fig.~\ref{fig:1}. The form factor can be accurately fitted by
\begin{equation}\label{eq:para}
f_+(q^2) = \frac{f_+(0)}{\displaystyle 1 - a\,(q^2/m_B^2) +
    b\,(q^2/m_B^2)^2}\,,
\end{equation}
with $f_+(0)$, $a$ and $b$ given in Tab.~1, for different values of
$m_b$, $s_0$ and $M^2$.
The above parametrization reproduces the actual values calculated from the
LCSR, for $q^2\leq 14\,$GeV$^2$, to within 2\% accuracy.

\begin{figure}[tbh]
\centerline{\epsfxsize=0.5\textwidth\epsffile{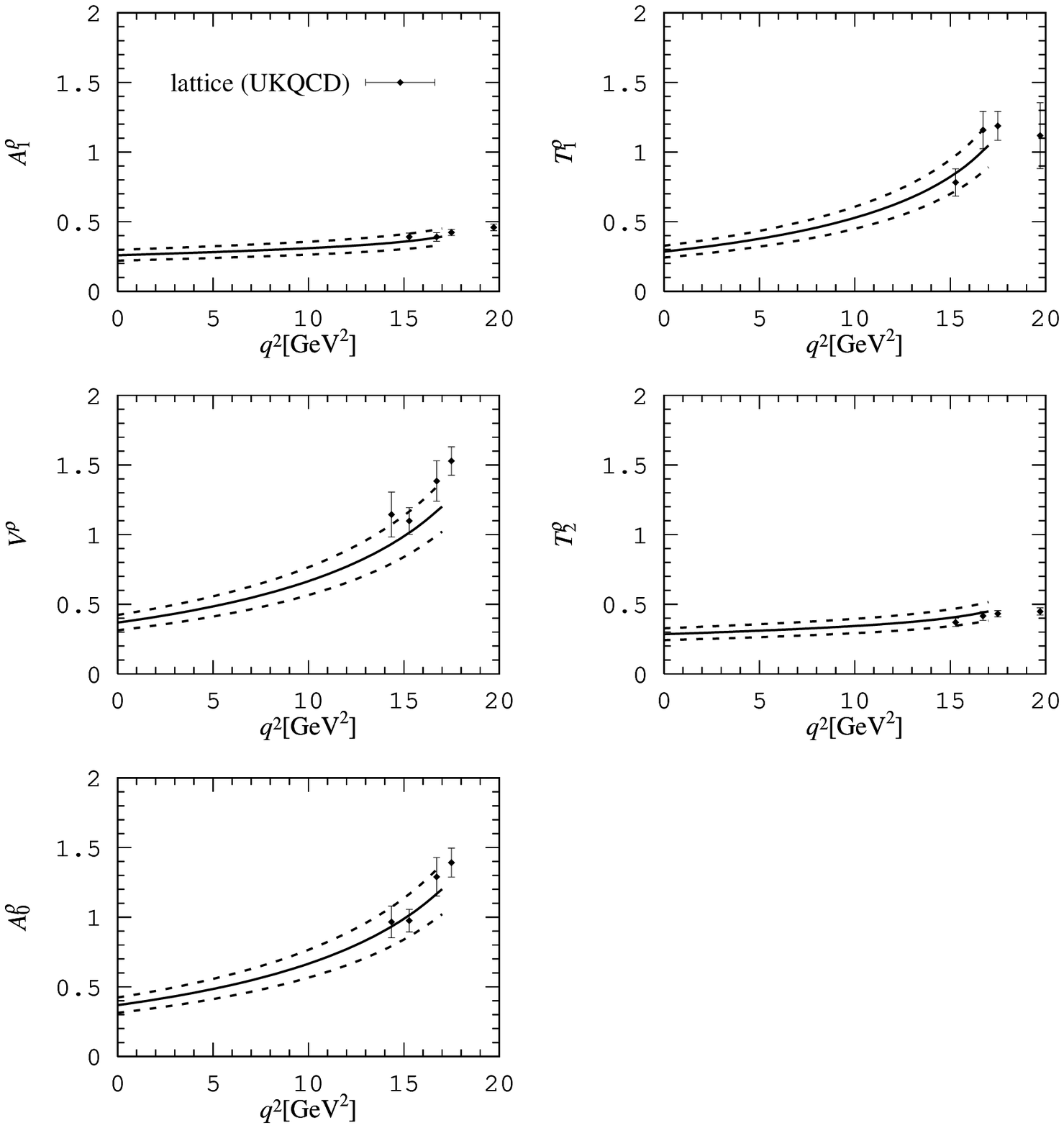}}
\vskip-10pt
\caption[]{Comparison of the light-cone sum-rule predictions
for the $B\to\rho$ form factors with lattice
calculations. Lattice errors
are statistical only.
The dashed curves show a 15\% uncertainty range of the sum rules
results. Figure taken from Ref.~\cite{BB98}.
}\label{fig:2}
\vskip10pt
\centerline{\epsfxsize=0.5\textwidth\epsffile{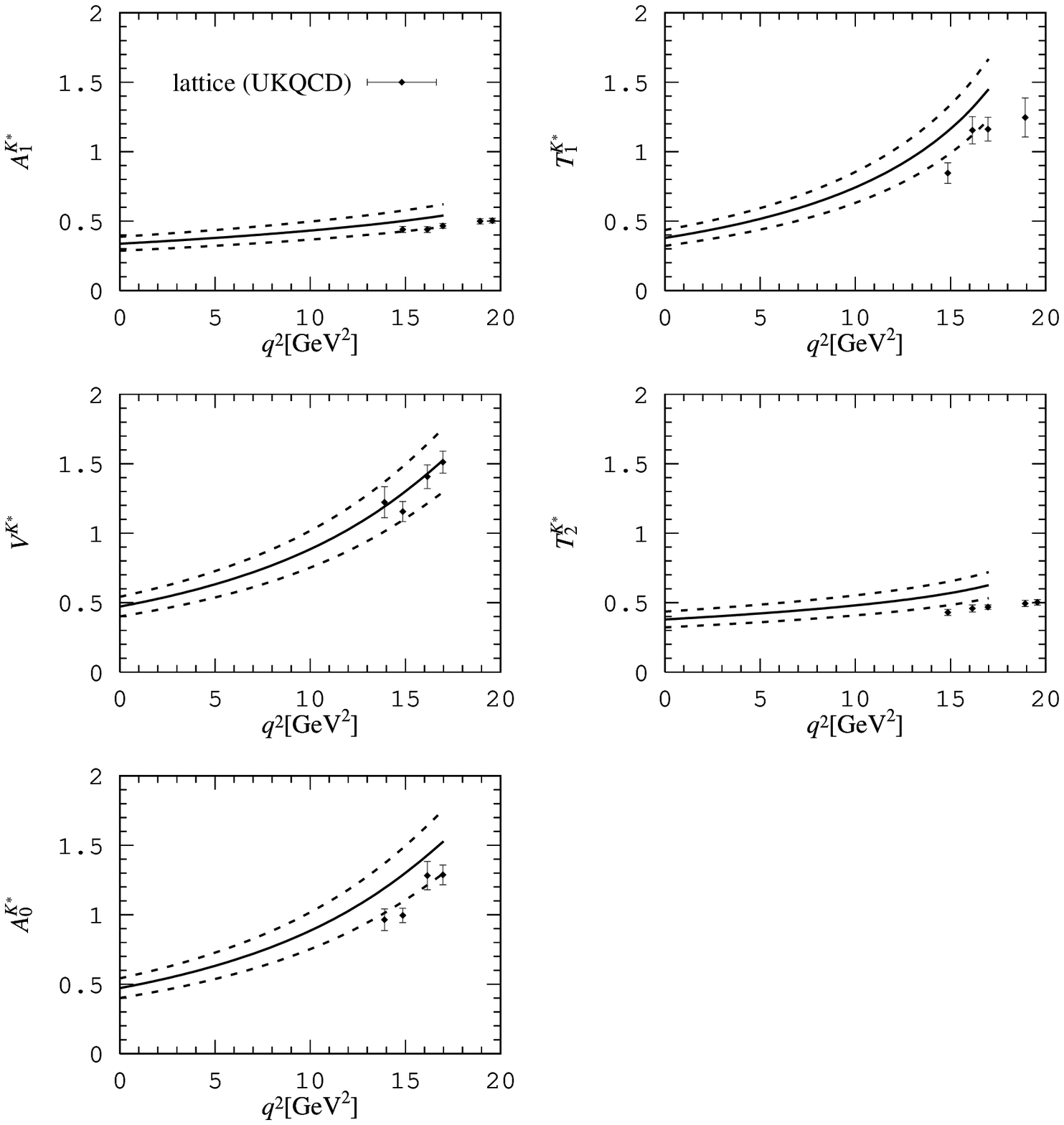}}
\vskip-10pt
\caption[]{Like Fig.~\ref{fig:2}, but for $B\to K^*$ form factors.
}\label{fig:3}
\end{figure}

Within the same method, also all the other form factors defined in
Eqs.~(\ref{FF1}) to (\ref{FF2}) can be calculated. In
Figs.~\ref{fig:2} and \ref{fig:3}
we plot the results for $B\to\rho$ and $B\to K^*$ form factors and
compare them with lattice calculations, which are available for large
$q^2$ only. It turns out that all form factors are very well described
by the three-parameter formula (\ref{eq:para}); we list the
corresponding best fit values in Tabs.~1 to 3.

\begin{table}[tbh]
{\small
\addtolength{\arraycolsep}{1pt}
$$
\begin{array}{|l|rrr|rrr|l|}
\hline
& \multicolumn{1}{c}{F(0)} &
\multicolumn{1}{c}{a_F} &
\multicolumn{1}{c|}{b_F} &
\multicolumn{1}{c}{F(0)} &
\multicolumn{1}{c}{a_F} &
\multicolumn{1}{c|}{b_F} &\\\hline
f_+^\pi & 0.261 & 2.03 & 1.293 & 0.341 & 1.41 & 0.406 & f_+^K\\
f_0^\pi & \equiv f_+^\pi(0) & 0.27 & -0.752 & \equiv f_+^K(0) & 0.41
& -0.361 & f_0^K\\
f_T^\pi & 0.296 & 1.28 & 0.193 & 0.374 & 1.42 & 0.434 & f_T^K\\
\hline
\end{array}
$$
\addtolength{\arraycolsep}{-1pt}
}
\caption[]{Results for $B\to P$ form factors for central values of the
  input parameters in the three-parameter fit of
  Eq.~(\ref{eq:para}). $f_T$ is renormalized at $\mu=m_b$.
The theoretical uncertainty is $\sim
  15\%$. Numbers from Refs.~\cite{solo,zwicky}.}
\renewcommand{\arraystretch}{1.3}
\addtolength{\arraycolsep}{3pt}
%\renewcommand{\arraystretch}{1.5}
%\addtolength{\arraycolsep}{3pt}
{\small\addtolength{\arraycolsep}{-3pt}
$$
\begin{array}{|l|ccc|ccc|l|}
\hline
& F(0) & a_F & b_F & F(0) & a_F & b_F & \\ \hline
A_1^{\rho} & 0.261 & 0.29 & -0.415& 0.337& 0.60 & -0.023& A_1^{K^*} \\
A_2^{\rho} & 0.223 & 0.93 & -0.092 & 0.283 & 1.18 & \phantom{-}0.281
& A_2^{K^*}\\
A_0^\rho & 0.372 & 1.40 & \phantom{-}0.437 & 0.470 & 1.55 &
\phantom{-}0.680 & A_0^{K^*} \\
V^\rho & 0.338 & 1.37 & \phantom{-}0.315
 & 0.458 & 1.55 & \phantom{-}0.575 & V^{K^*} \\ \hline
T_1^\rho & 0.285 & 1.41 & \phantom{-}0.361
 & 0.379 & 1.59 & \phantom{-}0.615 & T_1^{K^*}\\
T_2^\rho & 0.285 & 0.28 & -0.500
 & 0.379 & 0.49 & -0.241 & T_2^{K^*}\\
T_3^\rho & 0.202 & 1.06 & -0.076
 & 0.261 & 1.20 & \phantom{-}0.098 & T_3^{K^*}\\
\hline
\end{array}
$$}
\caption{$B_{u,d}$ decay form factors in the three-parameter fit of
  Eq.$\,$(\ref{eq:para}).
Renormalization scale
for $T_i$ is $\mu = m_b$. The theoretical uncertainty is
estimated as 15\%. From Ref.~\cite{BB98}.}\label{tab:fit}
{\small\addtolength{\arraycolsep}{-3pt}
$$
\begin{array}{|l|ccc|ccc|l|}
\hline
& F(0) & a_F & b_F & F(0) & a_F & b_F & \\ \hline
A_1^{K^*} & 0.190 & 1.02 & -0.037& 0.296 & 0.87 & -0.061 &
A_1^{\phi} \\
A_2^{K^*} & 0.164 & 1.77 & \phantom{-}0.729 & 0.255 & 1.55 &
\phantom{-}0.513 & A_2^{\phi}\\
A_0^{K^*} & 0.254 & 1.87 & \phantom{-}0.887 & 0.382 & 1.77 &
\phantom{-}0.856 & A_0^{\phi} \\
V^{K^*} & 0.262 & 1.89 & \phantom{-}0.846
 & 0.433 & 1.75 & \phantom{-}0.736 & V^{\phi} \\ \hline
T_1^{K^*} & 0.219 & 1.93 & \phantom{-}0.904
 & 0.348 & 1.82 & \phantom{-}0.825 & T_1^{\phi}\\
T_2^{K^*} & 0.219 & 0.85 & -0.271
 & 0.348 & 0.70 & -0.315 & T_2^{\phi}\\
T_3^{K^*} & 0.161 & 1.69 & \phantom{-}0.579
 & 0.254 & 1.52 & \phantom{-}0.377 & T_3^{\phi}\\
\hline
\end{array}
$$
}
\caption{Like Tab.~\ref{tab:fit} for $B_s$ decays.}
\renewcommand{\arraystretch}{1}
\addtolength{\arraycolsep}{-3pt}
\end{table}

Progress in the accuracy of the LCSRs is possible, but
not likely to reduce the uncertainties dramatically. It would have to
come primarily from a reduction in the uncertainty of input
parameters, i.e.\ the meson decay constants and distribution
amplitudes, which could come either from lattice calculations 
or, in particular for
strange mesons, from a re-evaluation of SU(3) breaking effects from
QCD sum rules \cite{atwork}.

\end{document}